\documentclass[11pt,twoside]{article}
\usepackage{asp2010}
\aspvolume{\textsf{\rule{8mm}{0mm}}}
\aspcpryear{2012}
\aspvoltitle{Magnetic Fields from the Photosphere to the Corona,
    2$^\mathit{nd}$ ATST-EST Workshop}
\aspvolauthor{\rule{8mm}{0mm}}
\resetcounters

\usepackage{natbib}
\usepackage{amssymb}
\usepackage[usenames, dvipsnames]{color}

 \usepackage[T1]{fontenc}
 \usepackage{textcomp}
 \usepackage{mathptmx}
 \usepackage{courier}
 \usepackage[scaled=0.94]{helvet}
 \usepackage{hyperref}
 \hypersetup{
     final=true,
     pageanchor=true,
     hyperfigures=false,
     colorlinks=true,
     breaklinks=true,
     linkcolor=blue,
     citecolor=blue,
     urlcolor=blue,
     pdfpagemode=UseNone,
     pdfpagelayout=OneColumn,
     pdfview=FitBH,
     pdftitle={The GREGOR Fabry-P\'erot Interferometer - A New Instrument for  High-Resolution Spectropolarimetric Solar Observations},
     pdfauthor={Puschmann et al.},
     pdfsubject={Solar Physics}}

\begin{document}

%
%

\title{The GREGOR Fabry-P\'erot Interferometer -- A New Instrument for High-Resolution Spectropolarimetric Solar Observations}
\author{Klaus G. Puschmann,$^1$
    Horst Balthasar,$^1$
    Svend-Marian Bauer,$^1$\\
    Thomas Hahn,$^1$
    Emil Popow,$^1$
    Thomas Seelemann,$^2$
    Reiner Volkmer,$^3$\\
    Manfred Woche,$^1$ and
    Carsten Denker$^1$
\affil{$^1$Leibniz-Institut f\"ur Astrophysik Potsdam,
    14482 Potsdam,
    Germany\smallskip\\
$^2$LaVision,
    37081 G\"ottingen,
    Germany\smallskip\\
$^3$Kiepenheuer-Institut f\"ur Sonnenphysik,
    79104 Freiburg,
    Germany}}

%
%

\begin{abstract}
Fabry-P\'erot interferometers have advantages over slit spectrographs because
they allow fast narrow-band imaging and post-factum image reconstruction of
spectropolarimetric data. Temperature, plasma velocity, and magnetic field maps
can be derived from inversions of photospheric and chromospheric spectral lines,
thus, advancing our understanding of the dynamic Sun and its magnetic fields at
the smallest spatial scales. The GREGOR Fabry-P\'erot Interferometer (GFPI) is
one of two first-light instruments of the 1.5-meter GREGOR solar telescope,
which is currently being commissioned at the Observatorio del Teide, Tenerife,
Spain. The GFPI operates close to the diffraction limit of GREGOR, thus,
providing access to fine structures as small as 60~km on the solar surface. The
field-of-view of $52\arcsec \times 40\arcsec$ is sufficiently large to cover
significant portions of active regions. The GFPI is a tuneable dual-etalon system
in a collimated mounting. Equipped with a full-Stokes polarimeter, it records
spectropolarimetric data with a spectral resolution of $R \approx 250,000$ over
the wavelength range from 530--860~nm. Large-format, high-cadence CCD detectors
with powerful computer hard- and software facilitate scanning of spectral lines
in time spans corresponding to the evolution time-scale of solar features. We
present the main characteristics of the GFPI including the latest developments
in software, mechanical mounts, and optics.
\end{abstract}

%
%

\section{Introduction}

The determination of the thermodynamic and magnetic structure of the solar atmosphere at the smallest spatial scales is one of the cornerstones of modern solar physics. An accurate derivation of the atmospheric properties requires both high spatial and spectral resolution. Nowadays, almost all ground-based solar telescopes are equipped with FPIs. A summary of available instruments was
recently provided by \citet{puschmannbeck11}. In 1986, the
Universit\"ats-Sternwarte G\"ottingen developed an imaging spectrometer for the German Vacuum Tower Telescope. This instrument used a universal birefringent filter (UBF) as a pre-filter for order sorting of a narrow-band FPI \citep{bendlinetal92}. The spectrometer was later equipped with a Stokes-$V$ polarimeter \citep{volkmer+etal95}. The UBF was replaced in 2000 by a second etalon. A fundamental renewal of the G\"ottingen FPI began during the first half of 2005 \citep{puschmann+etal2006} in preparation
for the 1.5-meter GREGOR solar telescope \citep{volkmer+etal2010a,
volkmer+etal2010b}. New narrow-band etalons were integrated and new large-format, high-cadence CCD detectors with powerful computer hard- and software provide now the means for fast data acquisition and user-friendly operation. The instrument was also upgraded to a full-Stokes spectropolarimeter \citep{nazi+kneer2008a, balthasaretal09, balthasar+etal2011}. The optical design and instrument characteristics of the GREGOR Fabry-P\'erot
Interferometer (GFPI) were presented by \citet{puschmannetal07} and more recently by \citet{denker+etal2010}. The Leibniz-Institut f\"ur Astrophysik Potsdam took over the scientific responsability  for the GFPI in 2009, when the GFPI was transfered to the GREGOR solar telescope, where it is currently being commissioned.

%
%

\section{GREGOR Fabry-P\'erot Interferometer}

The GFPI data acquisition system consists of two Imager QE CCDs from LaVision, which have a Sony ICX285AL chip with a full well capacity of 18,000~e$^{-}$ and a read-out noise of 4.5~e$^{-}$. The analog-digital conversion is carried out with 12-bit resolution. The detectors have a spectral response from 320--900~nm
with a maximum quantum efficiency of 60\% at 550~nm. The detector has $1376 \times 1040$ pixels with a size of $6.45 \times 6.45$~$\mu$m$^{2}$ resulting in a total area of $8.8 \times 6.7$~mm$^{2}$. The two etalons manufactured by IC
Optical Systems (ICOS) have a diameter of $\oslash = 70$~mm, a finesse of ${\cal F} \sim 46$, and spacings of $d = 1.1$~mm and 1.4~mm. They are mounted near a pupil image in the collimated light beam and have a high-reflectivity coating ($R \sim 95$\%) in the wavelength range from 530--860~nm. A narrow transmission
curve with a $\mathrm{FWHM} = 1.9$--5.6~pm results in a high spectral purity and spectral resolution of ${\cal R} = 250.000$. All etalons are operated by three-channel CS100 controllers with a three-axis capacitance bridge stabilization system, which ensures the parallelism and cavity spacing of the etalon. The cavity spacing itself is digitally controlled via RS-232 communication with the GFPI control computer so that a spectral region can be rapidly scanned. The etalons are protected by a small box to keep them in a
thermally stable environment. The spectrometer is equipped with a full-Stokes polarimeter. The modulation is performed with two ferro-electric liquid crystals -- one acting as quarter-wave plate and the other as half-wave plate. A modified Savart plate serves as polarimetric beam splitter. The polarimeter operates in
the wavelength range between 580--660~nm with an optimal performance at 630~nm.

\begin{figure}[t]
\centerline{\includegraphics[width=0.88\textwidth]{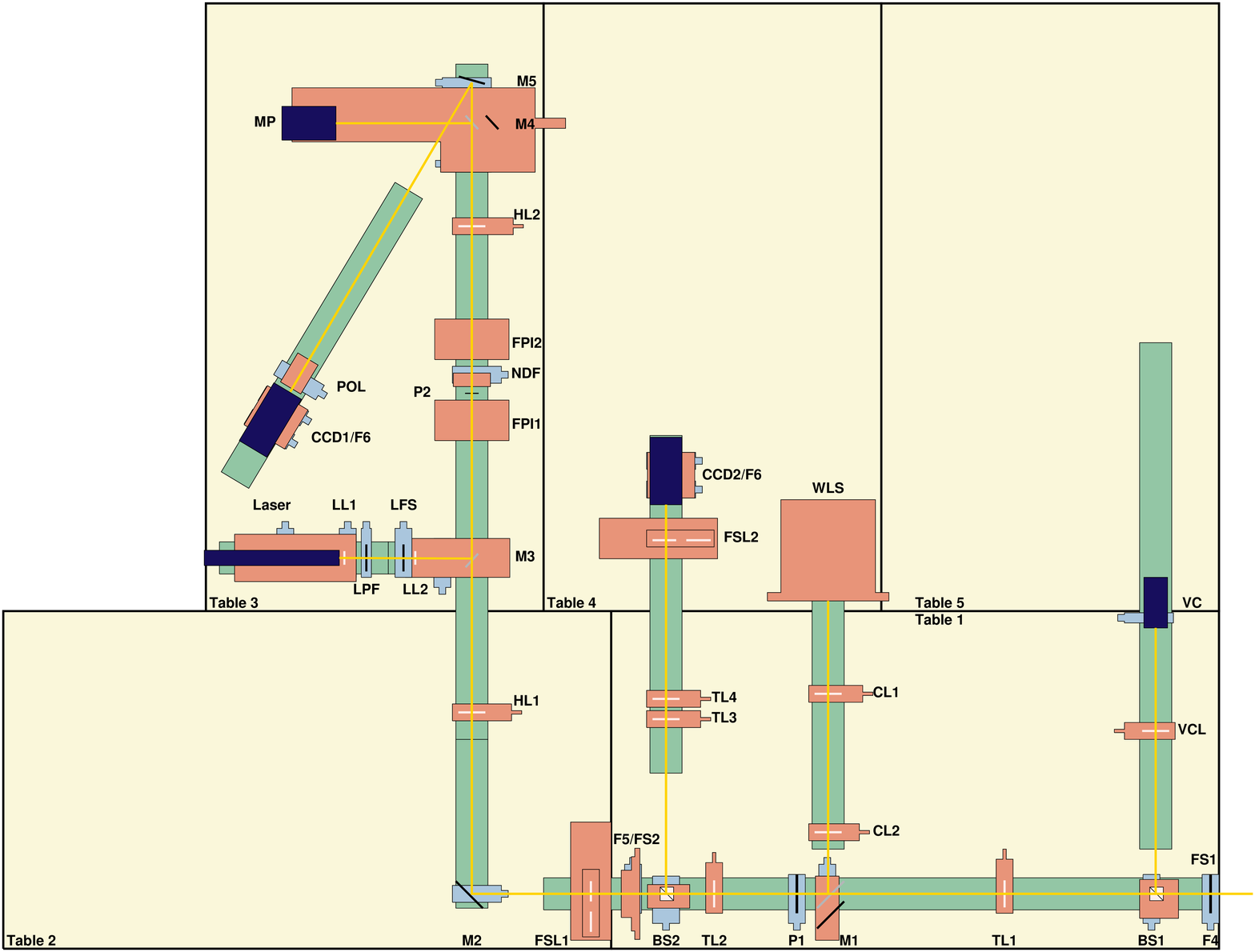}}
\caption{
    \textsf{CCD1} \& \textsf{CCD2}: CCD detectors,
    \textsf{FPI1} \& \textsf{FPI2}: narrow-band etalons,
    \textsf{NDF}: neutral density filter,
    \textsf{TL1}, \textsf{TL2}, \textsf{TL3}, \textsf{TL4}, \textsf{HL1},
        \textsf{HL2}, \& \textsf{VCL}: achromatic lenses,
    \textsf{CL1} \& \textsf{CL2}: plano-convex lenses,
    \textsf{M1}, \textsf{M3}, \& \textsf{M4}: removable folding mirrors,
    \textsf{M2} \& \textsf{M5}: fixed folding mirrors,
    \textsf{F4}, \textsf{F5}, \& \textsf{F6}: foci,
    \textsf{P1} \& \textsf{P2}: pupil images,
    \textsf{BS1} \& \textsf{BS2}: beam-splitters,
    \textsf{FS1} \& \textsf{FS2}: field stops,
    \textsf{VC}: video camera,
    \textsf{WLS}: white-light source (slide projector),
    \textsf{FSL1} \& \textsf{FSL2}: filter sliders,
    \textsf{POL}: full-Stokes polarimeter,
    \textsf{LL1} \& \textsf{LL2}: laser lenses,
    \textsf{LPF}: laser polarization filter,
    \textsf{LFS}: laser field stop, and
    \textsf{MP}: photomultiplier.}
\label{FIG01}
\end{figure}

The GFPI is located in an optical laboratory at the  5$^{\rm th}$ floor of the GREGOR building. It occupies five optical tables as shown in Fig.~\ref{FIG01}, where it is protected by an aluminum housing 80-cm tall, which safeguards the optics from dust contamination and prevents straylight from entering the optical path. 
A dichroic pentaprisms just in front of the telescope's science focus \textsf{F4} sends a small portion of the green continuum to the wavefront sensor of the adaptive optics (AO) system \citep{Berkefeld2010}.
Immediately, another dichroic pentaprisms with a
cut-off at H$\alpha$ $\lambda 656.28$~nm directs the light to the Grating Infrared Spectrograph \citep[GRIS,][]{Collados2010} and the GFPI so that
simultaneous observations in the infrared and visible wavelength regions are possible. Four achromatic lenses \textsf{TL1}, \textsf{TL2}, \textsf{HL1}, and \textsf{HL2} in the narrow-band optical channel \textsf{NBC} create two foci \textsf{F5} \& \textsf{F6} and two pupil images \textsf{P1} \& \textsf{P2}. The etalons
\textsf{FPI1} \& \textsf{FPI2} are placed in the vicinity of P2. A neutral
density filter \textsf{NDF} with a transmission of 63\% between the etalons reduces inter-etalon reflexes. In the \textsf{NBC} the light beam is folded twice by \textsf{M2} and \textsf{M5}. The field stops \textsf{FS1} \& \textsf{FS2} prevent straylight, \textsf{FS2} avoids an overlapping of the images separated by the full-Stokes polarimeter \textsf{POL}, which can be placed in front of
\textsf{CCD1}. The beam splitting cubes \textsf{BS1} \& \textsf{BS2} near foci \textsf{F4} \& \textsf{F5} send the blue part of the spectrum (below 530~nm) to a video camera \textsf{VC} and just 5\% of the light to the broad-band channel \textsf{BBC}. The achromatic lenses \textsf{TL3} \& \textsf{TL4} are chosen such that the image scale in both \textsf{BBC} and \textsf{NBC} (\textsf{F6}) is exactly the same, i.e., 0\farcs038~pixel$^{-1}$ resulting in a
field-of-view (FOV) of $52\farcs2 \times 39\farcs5$. Interference filters with a FWHM of 5--10~nm and 0.3--0.8~nm restrict the bandpass for the BBC and NBC channels, respectively. Each channel is equipped with a computer-controlled filter slider (\textsf{FSL1}, \textsf{FSL2}) so that two spectral regions can be sequentially observed. Finally, a white-light channel for spectral calibration and a laser/photomultiplier channel for finesse adjustment of the etalons complete the optical set-up.

%
%

\section{Computer Hard- and Software and Device Communications}

The GFPI is operated by a control computer under Windows XP Professional, which has been recently upgraded \citep{denker+etal2010}. The motherboard is equipped
with an Intel Core~2 Quad CPU with a clock rate of 2.66~GHz and 3.0~GB RAM. Capturing two 12-bit images with a total of 2.6 megapixels at a data acquisition rate of 10~Hz, while simultaneously writing the data to disk, requires some special adaption of the hardware. Images of both CCD cameras are taken exactly
at the same time. Cameras are triggered by a Programmable Timing Unit (\textsf{PTU}) and the images are transferred to the computer as digital signals via two twin-coaxial cables, which are connected to two separate PCI boards. Image pairs of the narrow- and broad-band channels  are combined into one image. Images are first compressed on-the-fly using a zlib-based compression algorithm
before they are written to a 4~TB RAID0 disk array, which is controlled by an Adaptec 38-5 SCSI RAID controller. The system partition and additional temporary data storage are hosted on a separate 1~TB harddisk. Fast data transfer to the data storage and processing servers is enabled by two Gigabit network adapters
(Intel 82566DM-2 and Intel PRO/1000 PL).

The DaVis imaging software from LaVision in G\"ottingen (\url{www.lavision.de}) provides the means for operating the GFPI . In 2005,  DaVis~7.0 has been adapted to the needs of the spectrometer \citep{puschmann+etal2006}. The current version is DaVis~7.2, which has an improved graphical user interface (GUI), customized interfaces for hardware communication, and a more streamlined data
handling for high-speed imaging. Davis is based on a macro programming language (CL-language) with a syntax similar to C++ and provides fine-grained control for data acquisition, hardware components, user-defined observing procedures, and tools for quick-look data analysis. The modular software design makes it also straightforward to adapt the GFPI control to any future requirements, in particular with respect to systems integration and more complex observing sequences. Recently, three precision translation stages were added to automate observing procedures. Two stages are used to switch between two sets of interference filters in the
narrow- and broad-band channels. Thus, two separate wavelength bands can be sequentially observed. The third stage inserts a deflection mirror into the light path so that calibration data can be taken with a continuum light source.

The GUI consists of several menus: \textsf{Hardware Settings} $\rightarrow$ selection of cameras, binning, exposure time, frame rate, and FOV, \textsf{Camera Matching} $\rightarrow$ continuous camera read-out and superposition of the two camera images for pixel-accurate alignment, \textsf{Region of Interest} $\rightarrow$ selection of rectangular areas to visualize properties of a scanned line profile, \textsf{FPI Adjustment} $\rightarrow$ individual etalon control, algorithms for etalon alignment, and precise line positioning, and \textsf{Observation} $\rightarrow$ selection of
main observing parameters: number of scans, cycle time, and scan tables. Scan tables are user-defined lists of instructions, which contain information about pre-filters, scan ranges, step widths in wavelength, and the number of images per wavelength position. During the observations, the pre-filter, image number, wavelength position, scan number, scan time, data processing time and total
cycle time are displayed in real-time. At the end of each scan, the recorded images and the corresponding average line profiles can be displayed. However, this option can be disabled to reduce the total cycle time. All parameters are collected in a project file, which together with the imaging data, creates a self-describing data set.

\begin{figure}[t]
\centerline{\includegraphics[width=0.88\textwidth]{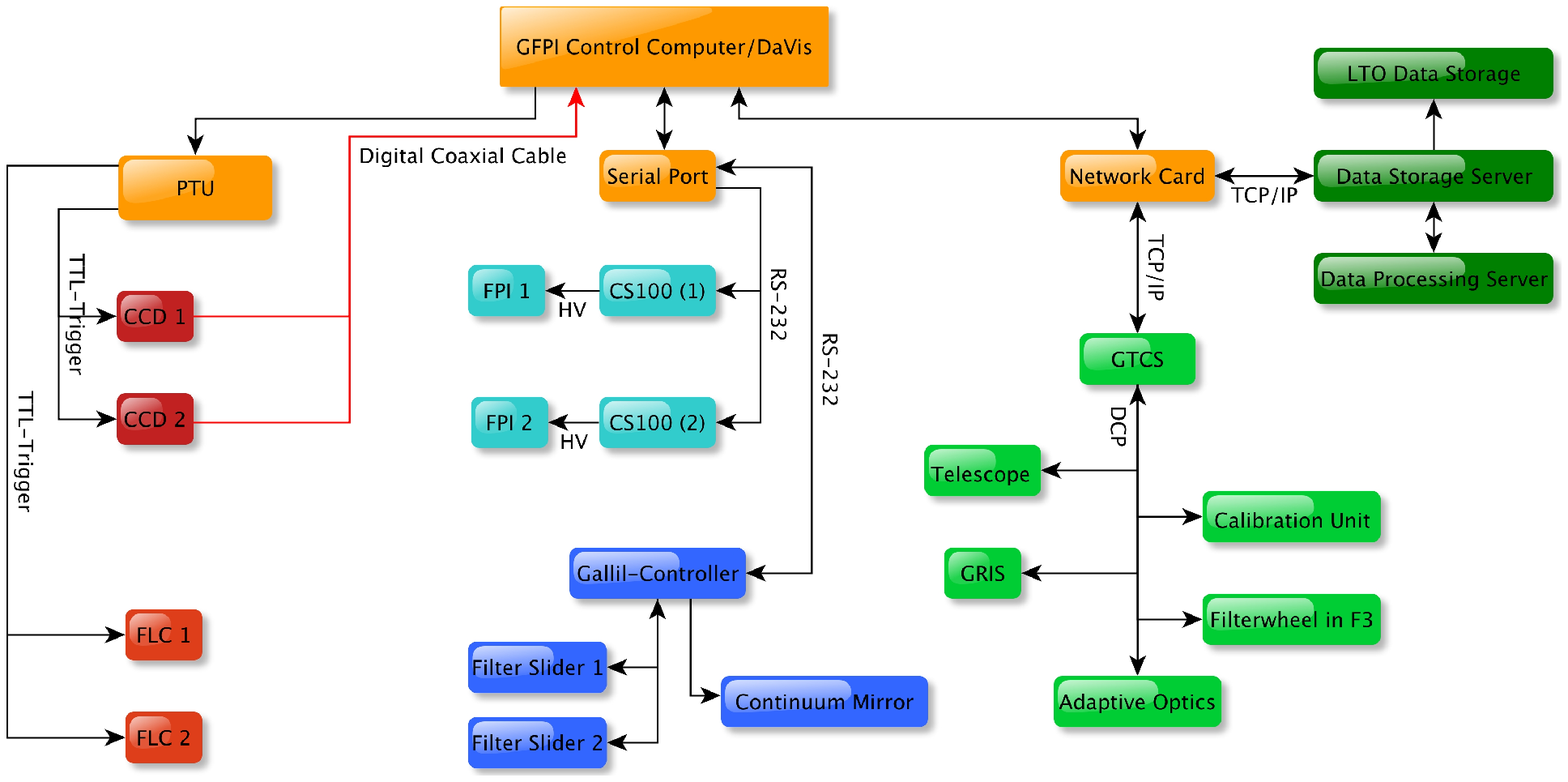}}
\caption{Flow chart of the communications between the GFPI control computer
    (DaVis) and internal and peripheral devices. \textsf{PTU}: programmable
    timing unit, \textsf{CCD~1} \& \textsf{CCD~2}: CCD detectors, \textsf{FLC~1}
    \& \textsf{FLC~2}: ferro-electric liquid crystals, \textsf{CS100 (1)} \&
    \textsf{CS100 (2)}: etalon controllers, and \textsf{FPI 1} \& \textsf{FPI
    2}: etalons.}
\label{FIG02}
\end{figure}

Fig.~\ref{FIG02} illustrates the communication between GFPI control computer and internal/external devices. Cameras \textsf{CCD~1} \& \textsf{CCD~2} and ferro-electric liquid crystals \textsf{FLC~1} \& \textsf{FLC~2} are triggered by a programmable timing unit \textsf{PTU} by analog TTL signals. The data recorded
by the cameras are passed via digital coaxial cables to two PCI cards inside the control computer. The control signals for the cavity spacings of the etalons \textsf{FPI 1} \& \textsf{FPI 2} are transmitted according to the RS-232 standard for serial communications. Similarly, RS-232 communications is used for
the precision translation stages, which are operated by an eight-axes DMC-2183 Galil controller. The communication with other external devices is implemented using TCP/IP, i.e., the GREGOR Telescope Control System (\textsf{GTCS}), polarimetric calibration unit, adaptive optics (AO), AO filter wheel, and GRIS.
The Device Communication Protocol (\textsf{DCP}) was specifically developed for exchanging messages/commands as ASCII strings. All DCP communications is handled by a central message relay server. Finally, the recorded scientific data can be transferred via TCP/IP to a data storage server, which saves them on a local
RAID-system so that it can be copied to LTO tapes for transport. Preliminary data reduction, quick-look data products, and conversion to standard FITS format using image extensions is performed by a separate data processing server. In the future, the software data pipeline on the data processing server will enable full processing of the spectral imaging data including image restoration.

%
%

\section{Data Processing and Image Restoration}

Imaging spectropolarimetry with Fabry-P\'erot interferometers has the advantage that post-factum image restoration can be applied improving spatial resolution across the entire FOV, thus, augmenting the real-time correction of AO systems. Two techniques have been successfully used with data from imaging spectropolarimetry: (1) speckle reconstruction with subsequent narrow-band
deconvolution and (2) blind deconvolution. At the moment, the GFPI data pipeline includes as one option the G\"ottingen speckle reconstruction code \citep{deboer93}, which was later improved by \citet{puschmann+sailer2006} and \citet{denker+etal2007} to account for the field-dependent real-time AO correction. Broad-band images can be directly speckle reconstructed because of a sufficient number of images with a high signal-to-noise ratio. Simultaneously recorded narrow-band images (always only a few images per wavelength position) are deconvolved in a second step using the speckle deconvolution method of
\citet{keller+vdluehe1992} by estimating the instantaneous object transfer function (OTF) from the observed and reconstructed broad-band images \citep{kriegetal99, janssen03}. Another implemented option is Multi-Frame Blind Deconvolution \citep[MFBD,][]{loefdahl2002}, which was extended by
\citet{vannortetal05} to cover also the case of multiple objects (MOMFBD), e.g., simultaneous images of the same object in slightly different wavelength settings or different polarization states. In the near future, the Kiepenheuer-Institute Speckle Interferometry Package \citep[KISIP,][]{woeger+etal2008, woeger+vdluehe2008} will be included in the GFPI data pipeline. KISIP pursues a
different approach accounting for the field-dependent real-time AO correction \citep{woeger2010} by using an analytic model of the AO system, which incorporates information from wavefront sensing. A comparison of speckle and blind deconvolution techniques was recently presented by \citet{puschmannbeck11}.

%
%

\acknowledgements The 1.5-meter GREGOR solar telescope is built and operated by the German consortium of the Kiepenheuer-In\-sti\-tut f\"ur Sonnenphysik in Freiburg, the Leibniz-In\-sti\-tut f\"ur Astrophysik Potsdam, and the Max-Planck-Gesellschaft in Munich with contributions by the Institut f\"ur Astrophysik G\"ottingen, the Instituto de Astrof{\'{\i}}sica de Canarias and other partners. CD acknowledges support by the Deutsche Forschungsgemeinschaft (DE~787/3-1).

%
%


\end{document}